\pdfoutput=1

\documentclass[12pt]{article}

\usepackage{amsfonts}
\usepackage{amsmath}
\usepackage{amssymb}
\usepackage{bigints}
\usepackage{booktabs}
\usepackage[nosort]{cite}
\usepackage{color}
\usepackage{dsfont}
\usepackage{float}
\usepackage{framed}
\usepackage{graphicx}
\usepackage{indentfirst}
\usepackage{mathrsfs}
\usepackage{multirow}
\usepackage{pdflscape}
\usepackage{setspace}
\usepackage{subdepth}
\usepackage{subfig}
\usepackage{titlesec}
\usepackage[dotinlabels]{titletoc}
\usepackage{wrapfig}
\usepackage[all]{xy}
\usepackage{young}
\usepackage[vcentermath]{youngtab}
\usepackage{relsize}
\usepackage{stackengine}
\usepackage{datetime}
\usepackage{physics}

\usepackage{hyperref}

\numberwithin{equation}{section}

\usepackage{verbatim}

\newcommand{\be}{\begin{equation}}
\newcommand{\ee}{\end{equation}}

\newcommand{\bml}{\begin{multline}}
\newcommand{\emll}{\end{multline}}

\newcommand{\nn}{\nonumber}

\def\({\left(} \def\){\right)}
\def\[{\left[} \def\]{\right]}

\def\mF{\mathcal{F}}

\def\al{\alpha}

\def\mO{\mathcal{O}}

\def\g{\gamma}

\def\lam{\lambda}

\def\d{\partial}

\newcommand{\la}{\langle}
\newcommand{\ra}{\rangle}

\newcommand{\bi}{\begin{itemize}}
\newcommand{\ei}{\end{itemize}}

\newcommand{\bea}{\begin{eqnarray}}
\newcommand{\eea}{\end{eqnarray}}

\usepackage[left=2cm,right=2cm,top=2cm,bottom=2cm]{geometry}
\titleformat{\section}{\large\bfseries}{\thesection.}{4pt}{}
\titlespacing{\section}{0pt}{22pt}{6pt}

\titleformat{\subsection}{\large\bfseries}{\thesubsection.}{4pt}{}
\titlespacing{\subsection}{0pt}{18pt}{6pt}

\titleformat{\subsubsection}{\normalfont\bfseries}{\thesubsubsection.}{4pt}{}
\titlespacing{\subsubsection}{0pt}{16pt}{6pt}

\def\ie{\begin{equation}\begin{aligned}}
\def\fe{\end{aligned}\end{equation}}

\def\tilde{\widetilde}
\def\t{\tilde}

\def\d{\partial}

\def\1{{\mathds 1}}

\def\mN{\mathcal{N}}

\DeclareFontShape{OT1}{cmr}{mx}{n}%
    {<->cmr10}{}
\newcommand{\mytitlefont}{\fontseries{mx}\selectfont}
\DeclareMathAlphabet{\titlemath}{OT1}{cmr}{mx}{n}

\def\sss{\subsubsection}

\begin{document}

\begin{titlepage}

\begin{center}

~\\[1cm]

{\fontsize{20pt}{0pt} \mytitlefont Correlation functions in linear chaotic maps}\\[10pt]

~\\[0.2cm]

{\fontsize{14pt}{0pt}Xu-Yao Hu{\small $^{1}$} and Vladimir Rosenhaus{\small $^{2}$}}

~\\[0.1cm]

 \it{$^1$Center for Cosmology and Particle Physics}\\ \it{ New York University} \\ \it{726 Broadway, New York, NY }\\[10pt]

\it{$^2$ Initiative for the Theoretical Sciences}\\ \it{ The  CUNY Graduate Center}\\ \it{
 365 Fifth Ave, New York, NY}\\[.5cm]

~\\[0.6cm]

\end{center}

\noindent 

The simplest examples of chaotic maps are linear, area-preserving maps on the circle, torus, or product of tori; respectively known  as the Bernoulli map, the cat map, and the recently introduced ``spatiotemporal'' cat map. We study correlation functions in these maps. For the Bernoulli map, we compute the correlation functions in a variety of ways: by direct computation of the integral, through Fourier series, through symbolic dynamics, and through periodic orbits. In relation to the more standard treatment in terms of eigenfunctions of the Perron-Frobenius operator, some of these methods are simpler and also extend to multipoint correlation functions. For the cat map, we compute correlation functions through a Fourier expansion, review and expand on a prior  treatment of two-point functions by Crawford and Cary, and discuss the limitations of shadowing. Finally, for the spatiotemporal cat map -- intended to be a model of many-body chaos -- we show that connected correlation functions of local operators vanish. 
\vfill

\newdateformat{UKvardate}{%
 \monthname[\THEMONTH]  \THEDAY,  \THEYEAR}
\UKvardate
\today\\
\end{titlepage}

\tableofcontents
~\\[.1cm]

\vspace{.5cm}

\section{Introduction}

At a microscopic level, chaotic systems are characterized by intricate phase space motion with rapidly diverging trajectories. Most of this detailed phase space structure is not, however, observable. What one can generally measure are quantities involving some kind of smearing in space or in time: the correlation functions. The correlation functions  in turn characterize, for instance, how rapidly the system approaches equilibrium and the transport coefficients. An important problem is computing correlation functions for chaotic systems.

Recently, \cite{GC} revisited some of the simplest chaotic maps, such as the Bernoulli map and the cat map, with a  view towards  understanding a many-body chaotic map, dubbed the ``spatiotemporal'' cat map \cite{Gutkin16}. These chaotic maps are linear, area-preserving maps on the circle (Bernoulli map), torus (cat map), and products of tori (spatiotemporal cat map), respectively. Gutkin and collaborators \cite{GC} studied the symbolic dynamics and periodic orbits of these maps, emphasizing that they should be viewed as discrete-time dynamical systems. In this paper we study the correlation functions in these maps, which can be computed due to the linearity of the maps. Our results complement and extend previous results in the literature on correlation functions for cat maps \cite{CrawfordCary}.

In Sec.~\ref{sec:2} we compute correlation functions for the Bernoulli map using a variety of  techniques. Expanding the functions whose correlation functions we are computing in terms of a Fourier series is the most general, and extends to the cat map. We also compute the correlation functions using symbolic dynamics, which is a particularly efficient method for higher-point correlation functions.  Finally, we compute correlation functions using the method of periodic orbits. In Sec.~\ref{sec:3} we start by computing the Ruelle resonances \cite{Ruelle1986} for the cat map, clarifying some discrepancies in the literature. We then compute correlation functions using Fourier series, and elaborate on a related method used previously in \cite{CrawfordCary}. We discuss the limitations of shadowing \cite{GC} for computing late-time correlation functions. In Sec.~\ref{sec:4} we turn to  the recently introduced spatiotemporal cat map \cite{Gutkin16, GC}. After solving the equations of motion, we  show that connected correlation functions of local fields vanish. We end with some comments in Sec.~\ref{sec:5}.

\section{Bernoulli map} \label{sec:2}
\begin{figure}
\centering
\includegraphics[width=2.7in]{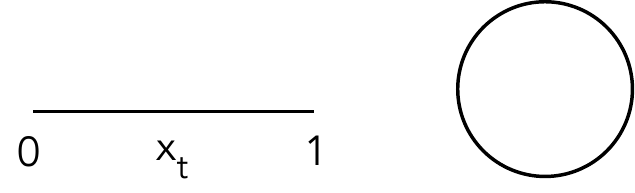}
\caption{The Bernoulli map acts on the interval $0\leq x<1$ with periodic identification (a circle). } \label{Bfig}
\end{figure}
The Bernoulli map is one of the simplest possible chaotic maps: at each integer time step $t$, the point $x_t$, which lives on a circle, is doubled: $x_{t+1} = 2 x_t \ \text{mod } 1$, see Fig.~\ref{Bfig}. We will write this as, 
\be
\delta_t x_t = x_t~ \ \ \ \text{mod } 1~, \ \ \ \ \ \delta_t x_t \equiv x_{t+1} - x_t~.
\ee
The solution is trivial, 
\be \label{xtb}
x_t = 2^t x_0~ \ \ \text{mod } 1~.
\ee 
In fact, discrete time is unnecessary. We may consider the continuous time map, $\d_t x(t) = \log 2\, x(t)$, where again $x(t)$ is confined to live on a circle. The solution is $x(t) = x_0 2^t$, so  (\ref{xtb}) is just viewing the system at discrete time intervals. \\

Let us now turn to the correlation functions \cite{Arnold}. In principle, we would like to start in some specific but generic state, evolve this state, and compute correlation functions in this state. As a result of ergodicity, the late-time correlation function of most local quantities will be the same if they are instead computed with the invariant measure. In the case of the Bernoulli map, this is the uniform distribution over $(0,1)$. So, for general functions $f(x)$ and $g(x)$, the two-point correlation function is defined as, 
\be \label{cdef}
\la f(x_t) g(x_0)\ra = \int_0^1 dx_0 \, f(x_t) g(x_0)~.
\ee
We will compute the two-point function using a variety of methods.~\footnote{See e.g.~\cite{Driebe} for an introduction to correlation functions in one-dimensional maps. For the Bernoulli map, \cite{Driebe} finds the two-point function  by solving for the eigenfunctions of the Perron-Frobenius operator. The method here is more direct.} 

\sss*{Direct computation}
Using the equations of motion, one can express $x_t$ in terms of $x_0$ (we assume $t$ is positive). 
For the Bernoulli map this is particularly simple, since we know the solution to the equations of motion, $x_t = 2^{t} x_0 \text{ mod } 1$. The nontrivial part here is the $\text{mod } 1$, which subtracts an integer. What we can do in (\ref{cdef}) is break the integral up into regions, depending  on the integer $n$ that is subtracted to implement the $\text{mod } 1$, 
\be
\la f(x_t) g(x_0)\ra = \sum_{n=0}^{2^t -1} \int_{n 2^{{-}t}}^{(n+1)2^{{-}t}} d x_0\, f(2^{t} x_0 {-} n) g(x_0)~.
\ee
For any particular $f(x)$ and $g(x)$, this integral and sum can be explicitly evaluated. 
For instance, 
\be \label{311c}
\la x_t x_0\ra = \frac{1}{4} + \frac{1}{12}\frac{1}{2^t}~, \ \ \ \ \la x_t^2 x_0^2\ra = \frac{1}{9} + \frac{1}{12}\frac{1}{2^t} + \frac{1}{180}\frac{1}{2^{2t}}~, \ \ \ \ \la x_t^3 x_0^3\ra = \frac{1}{16} + \frac{3}{40} \frac{1}{2^t}+ \frac{1}{80}\frac{1}{2^{2t}} - \frac{1}{140} \frac{1}{2^{3t}}~.
\ee
The correlation function $\la x_t^k x_0^k\ra$ will have terms that scales as $1/2^{ k t}$ and as lower powers. We can isolate the highest power by looking at correlation functions of the Bernoulli polynomials $B_k(x)$.  Doing this we find that for integer $k\geq 1$,
\be \label{Bcc}
\la B_k(x_t) B_k(x_0)\ra =\frac{(k!)^2 \zeta(2k)}{2^{2k{-}1}\pi^{2k}} \frac{1}{2^{kt}}~,
\ee
where $\zeta(x)$ is the Riemann zeta function, and $B_k(x)$ is a $k$'th order Bernoulli polynomial, whose expression we will soon give --  the first few are $B_1(x) = x{-}1/2$ and $B_2(x) = x^2 {-} x {+} 1/6$ and for e.g. $k=1,2$ (\ref{Bcc}) is, 
\be
\la B_1(x_t) B_1(x_0)\ra = \frac{1}{12}\frac{1}{2^t}~, \ \ \ \la B_2(x_t) B_2(x_0)\ra  = \frac{1}{180} \frac{1}{2^{2t}}~.
\ee
We can compute the correlation functions of any functions by expanding them in terms of the Bernoulli polynomials. 

\sss*{Fourier series}
We could instead compute the correlation functions through a Fourier series expansion. Because $x_t$ is defined with a $\text{mod } 1$, it should be the case that $f(x_t + 1) = f(x_t)$, so we can expand, 
\be
f(x_t) = \sum_{n=-\infty}^{\infty} f_n e^{2\pi i n x_t}~, \ \ \ \ g(x_0) = \sum_{m=-\infty}^{\infty} g_m e^{2\pi i m x_0}~.
\ee
In the exponential we may replace $x_t$ with $x_0 2^t$, as the $\text{mod } 1$ is now irrelevant. 
Now using the definition of the correlation function (\ref{cdef}), we have,~\footnote{The correlation function has time translation symmetry, i.e. $\ev{f(x_{t+\tau}) g(x_{\tau})} = \ev{f(x_{t}) g(x_{0})}$ for integer $\tau$, but not time reversal symmetry, $t\to -t$, because the Bernoulli map is not invertible.}
\be \label{fgF}
\la f(x_t) g(x_0)\ra  = \sum_{n,m = -\infty}^{\infty} f_n g_m \int_0^1 d x_0\,  e^{2\pi i (2^t n + m)x_0} = \sum_{n= -\infty}^{\infty} f_n\, g_{- 2^t n}~.
\ee

This result has a clear physical interpretation: as time evolves, an initial mode of wavenumber $n$ becomes a mode of effective wavenumber $2^t n$, 
\be
e^{2\pi i n  x_t} = e^{2\pi i ( 2^t n) x_0 }~.
\ee
This is of course a result of chaos, which involves  stretching and folding in phase space: the time evolution of $x_t$ is $x_t = 2^t x_0 \text{ mod } 1$, where the $2^t$ part encapsulates the stretching aspect of chaos, and the $\text{mod } 1$ implements the folding. For the correlator  $\la f(x_t) g(x_0)\ra$ to be nonzero at large $t$,  it has to be the case that $g(x)$ has some support in very high wavenumber modes (i.e. $d_{-2^t n}$ needs to be nonzero);  because, as  time evolves, those modes will get stretched and they need to end up having the same size as some of the modes making up $f(x)$. 
This gives a geometric way of seeing why correlation functions (of reasonably smooth functions) in chaotic systems rapidly decay with time: because, if the function is reasonably smooth, in Fourier space it will not have significant support for very  high wavenumber modes.

We may use the result (\ref{fgF}) to reproduce the correlation function of two Bernoulli polynomials. 
The Bernoulli polynomials have the Fourier series, 
\be \label{39}
B_k(x) = \frac{- k!}{(2\pi i )^k } \sum_{n\neq 0} \frac{e^{2\pi i n x}}{n^k}~.
\ee
Inserting this $f_n = \frac{- k!}{(2\pi i )^k }\frac{1}{n^k}$ into (\ref{fgF}) gives for the correlator of two Bernoulli polynomials, 
\be
\la B_k(x_t) B_k(x_0)\ra  = \(\frac{k!}{(2\pi i)^k}\)^2 \sum_{n \neq 0} \frac{1}{n^k}\frac{1}{(-2^t n)^k} =\frac{(k!)^2 }{2^{2k-1}\pi^{2k}} \frac{\zeta(2k)}{2^{kt}}~,
\ee
reproducing the answer (\ref{Bcc}) from before. 

\sss*{Symbolic dynamics}
The Bernoulli map involves, at each time step, doubling $x_t$ and then taking $\text{mod } 1$, to bring $x$ back within the range between $0$ and $1$. We may write the Bernoulli map in a symbolic dynamics form,
\be
\delta_t x_t = x_t - m_{t+1}~,
\ee
where $m_{t+1}$ is either $0$ or $1$. One can view the symbol $m_{t+1}$ as a source for a linear map\cite{GC}. The equations of motion can be solved in terms of a Green's function, 
\be\label{xt}
x_t = \sum_{t' = - \infty}^{\infty} g_{t t'} m_{t'}~, \ \ \ \ \ \ g_{t t'} = \frac{1}{2^{t'-t}} \theta(t'>t)~,
\ee
where $\theta(x)$ is the Heavside step function and $g_{t t'}$ is the Green's function satisfying $\delta_t g_{t, t'} - g_{t,t'} = -\delta_{t+1, t'}$ and is found by inspection.~\footnote{Explicitly, $g_{t+1, t'} = 2 g_{t t'}$ for $t' \neq t+1$, and $g_{t+1, t'}=0$ for $t' = t+1$. Thus  $\delta_t g_{t, t'} - g_{t,t'}\equiv g_{t+1,t'} - 2g_{t,t'}$ is equal to $0$ for $t'\neq t+1$, and is equal $1$ for $t' = t+1$.} This is of course just the standard Dyadic expansion familiar to the Bernoulli map, $x_t =  \sum_{t'=1}^{\infty} \frac{m_{t+t'}}{2^{t'}}$. 
Because of time-translation invariance, the Green's function only depends on the time difference, $
g_{t t'} = g(t-t')
$. 

The utility of this form of $x_t$ is that correlation functions of $x_t$ can be expressed in terms of correlation functions of  $m_t$, 
\be \label{321}
\la x_{t_1} \cdots x_{t_k}\ra = \sum_{t_1',\ldots, t_k'} g(t_1 {-} t_1') \cdots g(t_k {-} t_k') \la m_{t_1'} \cdots m_{t_k'}\ra~.
\ee
Of course, this is only useful if we know the correlation functions of the $m_t$. For the Bernoulli map this is indeed the case (however, as we will see later, for the cat map the situation is more involved).  For the Bernoulli map the probability distribution for the $m_i$ factorizes
\be \label{Pmi}
P(m_1, m_2, m_3, \ldots) = p(m_1) p(m_2) p(m_3) \cdots~, \ \ \ p(0) = p(1) = \frac{1}{2}~,
\ee
so the correlation functions of the $m_t$ factorize, 
\be \label{mf}
\la m_{t_1'} m_{t_2'}\ra \equiv \sum_{m_{t_1'} m_{t_2'}} P(m_{t_1'}, m_{t_2'})m_{t_1'} m_{t_2'}= \la m_{t_1'} \ra \la m_{t_2'}\ra ~, \ \ \ \ t_1' \neq t_2', \ \ \ \ \ \ \ \ \la m_t\ra = \la m_t^2 \ra = \frac{1}{2}~.
\ee
Explicitly, the first few correlation functions are, 
\bea \label{mc}
\la m_{t_1} m_{t_2}\ra = \frac{1}{4} \( \delta_{t_1, t_2} {+} 1 \)~, \ \ \ \la m_{t_1} m_{t_2} m_{t_3}\ra = \frac{1}{8} \( 1 {+} \delta_{t_1,t_2} {+} \delta_{t_1,t_3} {+} \delta_{t_2,t_3}\)~,\\
\la m_{t_1} m_{t_2} m_{t_3} m_{t_4}\ra = \frac{1}{16} \( 1 {+} \( \delta_{t_1,t_2} {+} \text{perm.}\) {+} \(\delta_{t_1,t_2} \delta_{t_3,t_4}{ +} \text{perm.}\) {-}2 \delta_{t_1,t_2} \delta_{t_1,t_3} \delta_{t_1,t_4} \)~.
\eea
The two-point function immediately follows, 
\be
\la x_{t_1} x_{t_2}\ra= \sum_{t_1', t_2' = -\infty}^{\infty} g(t_1 {-}t_1') g(t_2{ -} t_2') \frac{1}{4} \( \delta_{t_1', t_2'} +1\)=  \frac{1}{4} + \frac{1}{12}\frac{1}{2^{|t_{12}|}}~, \ \ \ \ \ t_{i j} \equiv t_i - t_j~,
\ee
reproducing the result (\ref{311c}) found earlier. 

This method is particularly useful for computing higher-point correlation functions. For instance, the three-point function is found by using (\ref{321}) combined with $\la m_{t_1'} m_{t_2'} m_{t_3'}\ra$ given in (\ref{mc}) and performing the sums, to get,  
\begin{align} \label{1130}
	\ev{x_{t_1}x_{t_2}x_{t_3}} = \frac{1}{8}+\frac{1}{24}\left(\frac{1}{2^{|t_{23}|}}+\frac{1}{2^{|t_{13}|}}+\frac{1}{2^{|t_{12}|}}\right) \ .
\end{align}
Finally, we look at the four-point function. Using (\ref{321}) combined with the connected part of $\la m_{t_1'} m_{t_2'} m_{t_3'}\ra$ (the last term in (\ref{mc})), the connected part of the four-point function is, \footnote{The full four-point function is
$
\la x_1 x_2 x_3 x_4\ra   = \la x_1 \ra \la x_2\ra \la x_3 \ra \la x_4\ra + \( \la x_1 x_2\ra_c \la x_3\ra \la x_4\ra + \text{perm.}\) 
+ \( \la x_1 x_2 \ra \la x_3 x_4\ra_c + \text{perm.}\) + \la x_1 x_2 x_3 x_4\ra  _c
$
where we didn't include $\la x(t_1) x(t_2) x(t_3) \ra_c$ because it is zero.} 
\be
\la x_{t_1} x_{t_2} x_{t_3} x_{t_4}\ra_c =- \frac{1}{2^3} \sum_{t_i' = t_i + 1} \prod_{i=1}^4 \frac{1}{2^{t_i' - t_i}}\,\, \delta_{t_1', t_2'} \delta_{t_1', t_3'} \delta_{t_1', t_4'} 
= -\frac{1}{2^3} \frac{1}{15}\frac{1}{2^{t_{12}+ t_{13}+t_{14}}}~, \ \ \ t_1\geq t_2, t_3, t_4~,
\ee
and permutations thereof if $t_2, t_3$ or $t_4$ is the greatest. 
 All higher-point correlation functions can be easily computed in a similar manner.

\sss*{Ruelle resonances}
The Perron-Frobenius operator $U$ acts on the space of functions, evolving some density forward in time
\be 
\rho_{t+1}(x) \equiv U \rho_t(x) = \int d x' \, \delta(x - S(x'))\, \rho_t(x')~,
\ee 
where we denoted the action of the Bernoulli map by $S$: $x_{t+1} = S x_t$. 

All correlation functions are straightforward to compute if one knows the eigenfunctions and eigenvalues of the Frobenius operator.~\footnote{Finding approximate eigenvalues/eigenfunctions of the Frobenius operator is sometimes possible and useful for realistic nonlinear systems  \cite{Mezic}.}  It is well-known that the eigenfunctions of the Frobenius operator for the Bernoulli map are Bernoulli polynomials and the eigenvalues are $ n \log 2$ for integer $n$. The eigenvalues are the Ruelle resonances \cite{Ruelle1986}, which govern the decay rate of the two-point function. From our computation of the two-point functions we have seen that they decay as $2^{-n t} = e^{- t n\log 2}$, in agreement with the Ruelle resonances. In general, it is not possible to find the eigenfunctions of the Frobenius operator;  finding the eigenvalues -- at least approximately -- is a simpler task, which can be done through the method of periodic orbits. We do this now for the Bernoulli map.

The eigenvalues of the Frobenius operator $U$ are given by $\log z$ where $z$ are such that  $
\det(1 - z U)$ has a zero. The determinant can be written as,
\be \label{729}
\det(1 - z U) = \exp \Tr \log(1 - z U) = \exp\( - \sum_{n=1}^{\infty} \frac{z^n}{n} \Tr U^n\)~.
\ee
We have that $\Tr U^n$ is,
\be \label{330}
\Tr U^n = \int dx \delta \(x - S^n(x)\) = \sum_{ x| x = S^n(x)} \frac{1}{|\frac{\d S^n}{\d x} - 1|}~.
\ee
So we need to find all periodic points $x$ that have a period $n$. From working out the first few values of $n$, it becomes clear that $S^n$ has fixed points at $x = \frac{k}{2^n -1}$ for $k=0, 1, \ldots 2^n -1$, and the slope of the map is $\frac{\d S^n}{\d x} = 2^n$. Hence $
\Tr U^n =\frac{2^n}{2^n - 1} $ and correspondingly, 
\be
\det(1 - z U) = \exp\( - \sum_{n=1}^{\infty} \frac{z^n}{n} \frac{1}{1 - 2^{-n}} \) = \exp\( - \sum_{n=1}^{\infty}\frac{z^n}{n} \sum_{k=0}^{\infty} 2^{- n k}\) 
=\prod_{k=0}^{\infty} \(1 - \frac{z}{2^k}\)~,
\ee
where we exchanged the order of the $n$ and $k$ sums and performed the $n$ sum to get $ \log \big(1 - \frac{z}{2^k}\big)$. 
We see that the zeros are at $z= 2^k$, and so  the eigenvalues (Ruelle resonances)  are indeed $k \log 2$ for positive integer $k$.~\footnote{See~\cite{Srednicki} for a discussion of Ruelle resonances in the closely related baker's map, and the implications for the quantum baker's map.}

\sss*{Periodic orbits}
The definition of a correlation function of some observable $\mO(x_t(x_0))$ is, 
\be
\la \mO\ra = \mN^{-1} \int d x_0\, \mO(x_t(x_0))~, \ \ \ \ \mN = \int d x_0~,
\ee
where we have written $x_t$ as a function of the initial condition $x_0$. The observable is a function of $x_t$ at various times $t$; it could be, for instance, the two-point function,  $\mO = x_{t_1} x_{t_2}$. We have also included a normalization factor $\mN$, which for the Bernoulli map is of course $1$. For some general system, to compute $\la \mO\ra$ numerically, we might imagine discretizing $x_0$, and for each $x_0$ solving the equation of motion to get the orbit, finding $\mO$ along the orbit, and then taking a sum over $x_0$. There is a great deal of freedom in deciding which $x_0$ to pick. 

A good choice of $x_0$ are the points that lie on periodic orbits, of period $n$,~\footnote{ See  \cite{EckhardtGrossmann,Cvit1995, gutzwiller,ChaosBook} for a discussion of dynamical averages in terms of periodic orbits in chaotic systems, and \cite{Cvit1, Cvit2} for applications to turbulence. }
\be
\la \mO \ra = \text{lim}_{n\rightarrow \infty}\ \   \mN^{-1} \int d x_0\, \delta(x_0 - S^{n}(x_0)) \mO(x_t(x_0))~, \ \ \ \mN =  \int d x_0\, \delta(x_0 - S^{n}(x_0))~,
\ee
where, as in (\ref{330}), we denote the map by $S$: $x_{t+1} = S x_t$, and eventually we want to take the limit of infinite $n$. The delta function localizes the integral to periodic orbits of period $n$, and we pick up a Jacobian,
\be \label{334}
\la \mO\ra =\text{lim}_{n\rightarrow \infty} \ \ \mN^{-1} \sum_{ x_0| x_0= S^n(x_0)} \frac{1}{|\frac{\d S^n}{\d x} - 1|} \mO(x_t(x_0))~, \ \ \ \ \ \ \mN = \sum_{ x_0| x_0 = S^n(x_0)} \frac{1}{|\frac{\d S^n}{\d x} - 1|}~.
\ee
As discussed below (\ref{330}), for the Bernoulli map, $S^n$ has fixed points at $x_0 = \frac{k}{2^n -1}$ for $k=0, 1, \ldots 2^n {-}1$. Thus, $\mN = 2^n/(2^n-1)$ and $\frac{\d S^n}{\d x} = 2^n$ and, 
\be \label{334v2}
\la \mO\ra =\text{lim}_{n\rightarrow \infty} \, 2^{-n}\!\! \sum_{ x_0| x_0= S^n(x_0)} \mO(x_t(x_0))~.
\ee
We now use this formula to compute the two-point function $\la x_{t_1} x_{t_2}\ra$. We represent $x_0$ in terms of the Dyadic expansion (\ref{xt}), 
$
x_0 = \sum_{k=1}^{\infty} \frac{m_k}{2^k}$. 
Periodic orbits of period $n$ correspond to those choices of $\{m_k\}$ for which $m_{k+n} = m_k$. We may use this periodicity to write $x_t$ as, 
\be \label{xrt}
x_t =\sum_{k=1}^{\infty}\frac{m_{k+t}}{2^k}= \frac{1}{1-2^{-n}}\sum_{k=1}^n\frac{m_{k+t}}{2^k}~,
\ee
where to get the second equality we used that the sum in each interval from $k= r n+1$ to $k=(r+1)n$  for integer $r$ is simply $2^{-n r}$ times the sum in the interval from $k=1$ to $k=n$. Further, we want to write $x_t$ so that it only involves $m_k$ with $k$ ranging from $1$ to $n$. Doing so, using the periodicity of $m_k$, gives, 
\be \label{337}
x_t = \frac{1}{1-2^{-n}}\( \frac{1}{2^n}\sum_{k=1}^{t} \frac{m_k}{2^{k-t}} +  \sum_{k=t+1}^{n} \frac{m_k}{2^{k-t}}\) =\frac{1}{1-2^{-n}} \sum_{k=1}^{n} \frac{m_k}{2^{k-t}}\(1 + (2^{-n}-1) \theta(k\leq t)\)~.
\ee
Applying (\ref{334v2}) gives us, 
\be \label{338}
\la x_{t_1} x_{t_2} \ra =\text{lim}_{n\rightarrow \infty}\, \, 2^{-n}\!\!\sum_{\{m_k\} | m_{k+n} = m_k} \!\!\!\!\! x_{t_1} x_{t_2}~,
\ee
where we are summing over all possible choices of $\{m_k\}$ which have periodicity $n$, of which there are $2^n$ such choices (as each $m_k$ is either $0$ or $1$). Inserting (\ref{337}) into (\ref{338}), 
\bea
\la x_{t_1} x_{t_2} \ra =\text{lim}_{n\rightarrow \infty}\, \, \frac{2^{-n}}{(1-2^{-n})^2}\!\!\sum_{\{m_k\} | m_{k+n} = m_k} \sum_{k_1, k_2=1}^n \frac{m_{k_1}}{2^{k_1-t_1}} \frac{m_{k_2}}{2^{k_2-t_2}}\(1 + (2^{-n}-1) \theta(k_1\leq t_1)\)\\ \nn
\(1 + (2^{-n}-1) \theta(k_2\leq t_2)\)
\eea
We split the sum over $k_1, k_2$ into a sum over terms with $k_1\neq k_2$ and those with $k_1 = k_2$, and then perform the sum over the $m_k$, 
\bea \nn
\la x_{t_1} x_{t_2} \ra =\!\text{lim}_{n\rightarrow \infty}\, \, \frac{2^{-n}}{(1{-}2^{-n})^2} \Big( 2^{n-2}\!\!\!\sum_{k_1 \neq k_2=1}^n \frac{1}{2^{k_1+k_2-t_1-t_2}}\[1 {+} (2^{-n}{-}1) \theta(k_1\leq t_1)\]
\[1 {+} (2^{-n}{-}1) \theta(k_2\leq t_2)\]  \\ \nn +2^{n-1} \sum_{k_1=1}^n \frac{1}{2^{2 k_1-t_1-t_2}}\[1 {+} (2^{-n}{-}1) \theta(k_1\leq t_1)\]
\[1 {+} (2^{-n}{-}1) \theta(k_1\leq t_2)\] 
\Big)
\eea
On the first line we got a factor of $2^{n-2}$, from summing over $m_k$ for $k\neq k_1, k_2$, whereas the $m_{k_1}$ and $m_{k_2}$ have to equal $1$, in order to give a nonzero contribution. Likewise, for the second line we got a factor of $2^{n-1}$  from summing over $m_k$ for $k\neq k_1$, whereas  $m_{k_1}$ has to equal one. 
Since the term on the second line is just twice the term on the first line with $k_1=k_2$,  we regroup terms, to write a sum over all $k_1, k_2$ plus half of the second line. The sum over all $k_1, k_2$ factorizes 
\be
\sum_{k_1=1}^n \frac{1}{2^{k_1-t_1}}\[1 + (2^{-n}-1) \theta(k_1\leq t_1)\]\sum_{k_2=1}^n  \frac{1}{2^{k_2-t_2}} \[1 + (2^{-n}-1) \theta(k_2\leq t_2)\]  = (1-2^{-n})^2~,
\ee
whereas the sum over $k_1$ gives, 
\be
\sum_{k_1=1}^n \frac{1}{2^{2 k_1-t_1-t_2}}\[1 {+} (2^{-n}-1) \theta(k_1\leq t_1)\]
\[1 {+} (2^{-n}-1) \theta(k_1\leq t_2)\]  = \frac{1}{3} (1-2^{-n}) \(\frac{1}{2^{|t_{12}|}} +  \frac{1}{2^{n-|t_{12}|}}\)~.
\ee
Thus, we finally get, 
\be
\la x_{t_1} x_{t_2} \ra =\text{lim}_{n\rightarrow \infty}\, \( \frac{1}{4} + \frac{1}{12} \frac{1}{2^{|t_{12}|}} \frac{1}{1-2^{-n}}+ \frac{1}{12} \frac{1}{2^{n-|t_{12}|}} \frac{1}{1-2^{-n}}\) = \frac{1}{4} + \frac{1}{12} \frac{1}{2^{|t_{12}|}}~.
\ee
Before taking the infinite $n$ limit, the answer has the symmetry $t_{12} \rightarrow n- |t_{12}|$, consistent with having periodicity $n$. After taking the infinite $n$ limit the result of course reproduces what we found earlier.

\section{Cat map} \label{sec:3}

\begin{figure}
\centering
\includegraphics[width=2.7in]{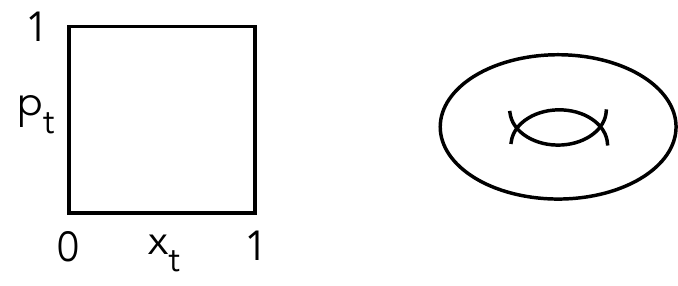}
\caption{The cat map acts on the two-dimensional phase space $(x,p)$ confined to a square with periodic identification (a torus). } \label{Cfig}
\end{figure}

The cat map  is a map on the torus $T^2$ and is commonly written as a map on two-dimensional $(x,p)$ phase space, with $x$ and $p$ running from $0$ to $1$  \cite{Arnold, GC,HannayBerry1980,Keating_1991}, 
\be \label{1620}
\binom{x_{t+1}}{p_{t+1}} = \(\begin{array}{cc} s{-}1 & 1 \\ s{-}2 & 1 \end{array}\) \binom{x_t}{p_t}~  \ \text{mod  } 1~,
\ee
where $s$ is an integer; for the Arnold cat map, $s=3$; see Fig.~\ref{Cfig}. The matrix has determinant equal to one, so it is an area preserving map. 
Following \cite{GC}, we let $p_t = x_t - x_{t-1}$, and write this as a discrete-time differential equation, 
\be \label{24}
\delta_t^2 x_t = (s-2) x_t~ \  \text{mod } 1, \ \ \ \ \ \delta_t^2 x_t \equiv x_{t+1} - 2 x_t + x_{t-1}~.
\ee
The solution is found by inserting the ansatz that $x_t$ is exponential in time, to get, 
\be \label{lameq}
x_t =c \lam^t + d \lam^{-t}~ \ \ \text{mod 1}~,\ \ \ \ \  \lam + \frac{1}{\lam} =s \, \ \  \Rightarrow  \ \  \lam = \frac{s}{2} + \sqrt{\frac{s^2}{4} - 1}~,
\ee
where $c$ and $d$ are constants which are fixed in terms of initial conditions. In terms of $t=0$ and $t=1$, $x_0 = c + d$ and  $x_1 = c \lam + \frac{d}{\lam}$. Alternatively, in terms of $x_0$ and $p_0$~\footnote{To write $p_0$ in terms of $x_0$ and $x_1$ we rely on the equations of motion,  $x_{t+1} = (s-1) x_t + p_t$, so 
at $t=0$, $p_0 = x_1 - (s-1)x_0 $. 
}
\bea \label{catx}
x_t &=&\frac{x_0 }{1+ \lam} \( \lam^{t+1} +\lam^{-t}\) + \frac{p_0}{ 1- \lam^2}\lam \( - \lam^{t}+\lam^{-t}\)\\
p_t &=& x_0 \frac{1- \lam}{1+ \lam} \( -\lam^t + \lam^{-t}\) + \frac{p_0}{1+\lam} \(\lam^t + \lam^{1-t}\)~.
\eea
Like with the Bernoulli map, we could have considered the continuous-time differential equation, 
\be
\d_t^2 x_t =(\log \lam)^2\,  x_t~,
\ee
which has the solution $x(t) \sim \lam^{\pm t}$, and, when viewed at integer times $t$, matches what we find from the cat map. 

Notice from the form of the equations of motion that the cat map can be viewed as an inverted harmonic oscillator, with potential $- \frac{1}{2}(s{-}2)(x \text{ mod } 1)^2$. The unboundedness of the inverted potential $-x^2$ is regulated through the $\text{mod } 1$. The map is chaotic for $s>2$ (positive $\lam$), which is the case we focus on.

\sss*{Ruelle resonances}
We begin by computing the Ruelle resonances for the cat map.~\footnote{There appears to be some discrepancy in the literature as to if there are nontrivial Ruelle resonances for the cat map. In particular, \cite{Keller2007,Frederic2007} state that the Ruelle resonance spectrum for the unperturbed cat map consists of only 0's and 1's. On the other hand, \cite{AQS1997} finds nontrivial Ruelle resonances for the Arnold cat map ($s=3$) which are consistent with our results.} As in the Bernoulli map, we apply formula (\ref{729}). Extending (\ref{330}) to two-dimensions, the trace of the $n$'th power of the Frobernius operator is again given by the sum over periodic orbits of period $n$, i.e. fixed points of the mapping $S(x,p)^n$ where $S(x,p)$ is the action of the cat map, given by the matrix on the right-hand side of (\ref{1620}), with the corresponding Jacobian factor.  
\be
J \equiv \frac{\d S(x,p)}{\d(x,p)} =  \(\begin{array}{cc} s{-}1 & 1 \\ s{-}2 & 1 \end{array}\)~, \ \ \ \  \ \ |\det( J^n -I )| = \lam^n + \lam^{-n} - 2~,
\ee
with $\lam$ given by (\ref{lameq}). As a result, $\text{Tr } U^n $ is given by ,
\be \label{trUn}
\text{Tr } U^n = \frac{N_n}{\lam^n + \lam^{-n} - 2}~,
\ee
where we defined $N_n$ to be the number of periodic orbits of period $n$, which factors out because the Jacobian factor in the denominator is a constant and independent of the orbit. 

Finding $N_n$ requires some care. The cat map acts on the region in $(x,p)$ space which is $\[0,1\) \times \[0,1\)$, because it is defined with a $\text{mod } 1$, so that $x+1$ is identified with $x$ and $p+1$ is identified with $p$. If we count the number of periodic orbits of period $n$ in this domain we find $N_n =  \lam^n + \lam^{-n} - 2$. With this $N_n$, (\ref{trUn}) gives $\text{Tr } U^n =1$ and correspondingly we would have $\det(1-zU) = 1-z$. As this only vanishes at $z=1$, this   would mean that there are no Ruelle resonances. This is not a satisfactory answer, as we will soon explicitly see that correlation functions of many reasonable functions decay exponentially.~\footnote{However, correlation functions of localized and infinitely differentiable  functions do indeed decay super-exponentially in the cat map \cite{CrawfordCary}, as we will see.}

To resolve this, we change the domain of the cat map: we take it to act on $\[0,1\] \times \[0,1\]$. Explicitly, we take the cat map to be the piecewise map, 
\be
S(x,p) = \begin{cases}  (2 x+p, x+p)~,  \ \ \ \ \ \ \ \ \ \ \ & 0\leq 2x+p \leq 1 \\
(2x+p{-}1, x+p)~, & 1< 2x+p \leq 2 \ \ \ \& \ \  0\leq x+p \leq 1 \\
(2x+p{-}1, x+p{-}1)~,& 1< 2x+p \leq 2 \ \ \ \& \ \   1< x+p \leq 2 \\
(2x+p{-}2, x+p{-}1)~, & 2< 2x+p \leq 3~,
\end{cases}
\label{cat S}
\ee
where we took $s=3$ (the Arnold cat map) in order to not clutter the equation. This now has one additional fixed point, $(x,p) = (1,1)$, which is no longer identified with $(0,0)$ and should be counted separately. The number of periodic orbits of period $n$ is therefore $N_n = \lam^n + \lam^{-n} - 1$ and thus, 
\be
\Tr U^n = 1 + \frac{1}{ \lam^n + \lam^{-n} -2} = 1+ \frac{ \lam^{-n}}{(1-\lam^{-n})^2}=1+ \sum_{m=0}^{\infty} (m+1) \lam^{-n(m+1)}~.
\ee
Correspondingly the determinant is,  
\be
\det (1 - z U) =\exp\( -\sum_{m=0}^{\infty}  (m+1)\sum_{n=1}^{\infty} \frac{1}{n}\(\frac{z}{ \lam^{m+1}}\)^n\) 
= \prod_{m=0}^{\infty} \( 1- \frac{z}{\lam^{m+1}} \)^{m+1}~.
\ee
The zeros are at $z = \lam^{m+1}$, and so the Ruelle resonances are $m \log \lam$ for positive integer $m$.

\sss*{Fourier series}
The two-point function is defined as, 
\be \label{2ptcat}
\la f(x_t, p_t) g(x_0, p_0) \ra = \int_0^1 dx_0 \int_0^1 d p_0\,  f(x_t, p_t) g(x_0, p_0)~.
\ee
We compute this through a Fourier series expansion,
\be \label{fourier}
f(x_t, p_t) = \sum_{m,n = -\infty}^{\infty} f_{m, n}\, e^{2 \pi i (m x_t + n p_t)}~, \ \ \ g(x_0, p_0) = \sum_{a,b = -\infty}^{\infty} g_{a, b}\, e^{2 \pi i (a x_0 + b p_0)}~.
\ee
We insert this into (\ref{2ptcat}) and exchange the order of summation and integration, 
\be \label{48}
\la f(x_t, p_t) g(x_0, p_0) \ra =  \sum_{m,n, a, b = -\infty}^{\infty}  f_{m, n}g_{a, b} \int_0^1 d x_0 dp_0 \,  \exp\(2 \pi i (m x_t {+} n p_t {+} a x_0 {+} b p_0)\)~.
\ee
 In order to perform the integral, we want to make use of the solutions of the equations of motion (\ref{catx}) expressing $x_t$ and $p_t$ in terms of $x_0$ and $p_0$. These can be written as, 
\bea \nn
x_t &=& ( 2 d_t{-}d_{t-1} ) x_0 + d_t\, p_0 \ \ \ \ \   \ \text{mod  } 1\\
p_t &=& d_t\, x_0 + ( 2 d_{t-1}{-}d_{t-2}) p_0\  \ \  \ \text{mod  } 1~, \ \ \ \ d_t = \frac{\lam}{\lam^2-1}( \lam^t - \lam^{-t})~, \label{xpt}
\eea
where $\lam$ was given in terms of $s$ in (\ref{lameq}). While it is not manifest, $d_t$ is an integer. For instance, for the Arnold cat map ($s=3$), $\lam$ and the first several values of $d_t$ are, 
\be
\lam = \frac{3 + \sqrt{5}}{2}~, \ \ \ \ \ d_1 = 1~, \ \ d_2 = 3~, \ \ \ d_3= 8~, \ \ \ d_4 = 21~, \ \ \ d_5 = 55~.
\ee
For this case, $d_t$ is the $(2t)$-th Fibonacci number.

Inside the exponential in (\ref{48}), the $\text{mod } 1$ appearing in the solution (\ref{xpt}) for $x_t$ and $p_t$ is irrelevant. Consequently, the integral is trivial, 
\be
 \int_0^1 d x_0 dp_0 \,  \exp\(2 \pi i (m x_t {+} n p_t {+} a x_0 {+} b p_0)\) = \delta_{a, -m ( 2 d_t{-}d_{t-1} ) - n d_t} \delta_{b, -n( 2 d_{t-1}{-}d_{t-2}) - m d_t}~,
\ee
where $\delta_{n,m}$ is the Kronecker delta function. The two-point function (\ref{48}) is thus, 
\be \label{twopC}
\la f(x_t, p_t) g(x_0, p_0) \ra =  \sum_{m,n = -\infty}^{\infty}  f_{m, n}\, g_{a,b} \Big|_{\substack{\!\!a =  -m ( 2 d_t{-}d_{t-1} ) - n d_t\\ \  b=  -n( 2 d_{t-1}{-}d_{t-2}) - m d_t}}~.
\ee
This expression can be compared with the analogous expression (\ref{fgF}) for the Bernoulli map. Compared to the Bernoulli map,  the cat map has a two-dimensional phase space and correspondingly the two-point function (\ref{twopC}) has a double sum. Unless one picks a special form for the coefficients $ f_{m, n}$ and $ g_{a,b}$, this sum is difficult to explicitly evaluate.

For instance, suppose we take $f$ and $g$ to be products of the  Bernoulli polynomials, for both $x$ and $p$,  $f(x,p) = g(x,p) = B_k(x) B_q(p)$. The Fourier coefficients are then (see Eq.~\ref{39}), 
\be
f_{m,n} = g_{m, n}= \frac{k! q!}{(2\pi i)^{k+q}}\frac{1}{m^k n^q}~, \ \ \ m, n \neq 0~,
\ee
and are equal to zero if either $m$ or $n$ is zero. Inserting into (\ref{twopC}), the two-point function is thus, 
\be \label{2ptBB}
\la  B_k(x_t) B_q(p_t ) B_k(x_0) B_q(p_0)\ra = \frac{k!^2 q!^2}{(2\pi)^{2(k+q)}}\sum' \frac{1}{m^k n^q}\frac{1}{\(m ( 2 d_t{-}d_{t-1} ) + n d_t\)^k}\frac{1}{ \(n( 2 d_{t-1}{-}d_{t-2}) {+} m d_t\)^q}~,   
\ee
where the sum is over all integers $m$ and $n$ for which none of the denominators vanish. We do not know how to analytically perform this sum. However, since $d_t$ scales as $\lam^t$, we see that at late times the correlator scales as $\lam^{-(k+q)t}$, 
\be \label{415}
\la  B_k(x_t) B_q(p_t ) B_k(x_0) B_q(p_0)\ra \sim \exp\( - \g_{k +q} t\)~, \ \ \ \ \g_n = n \log \lam~,
\ee
so it decays with the $k{+}q$'th Ruelle resonances.

\sss*{An alternate Fourier basis}
\begin{figure}
\centering
\includegraphics[width=2.5in]{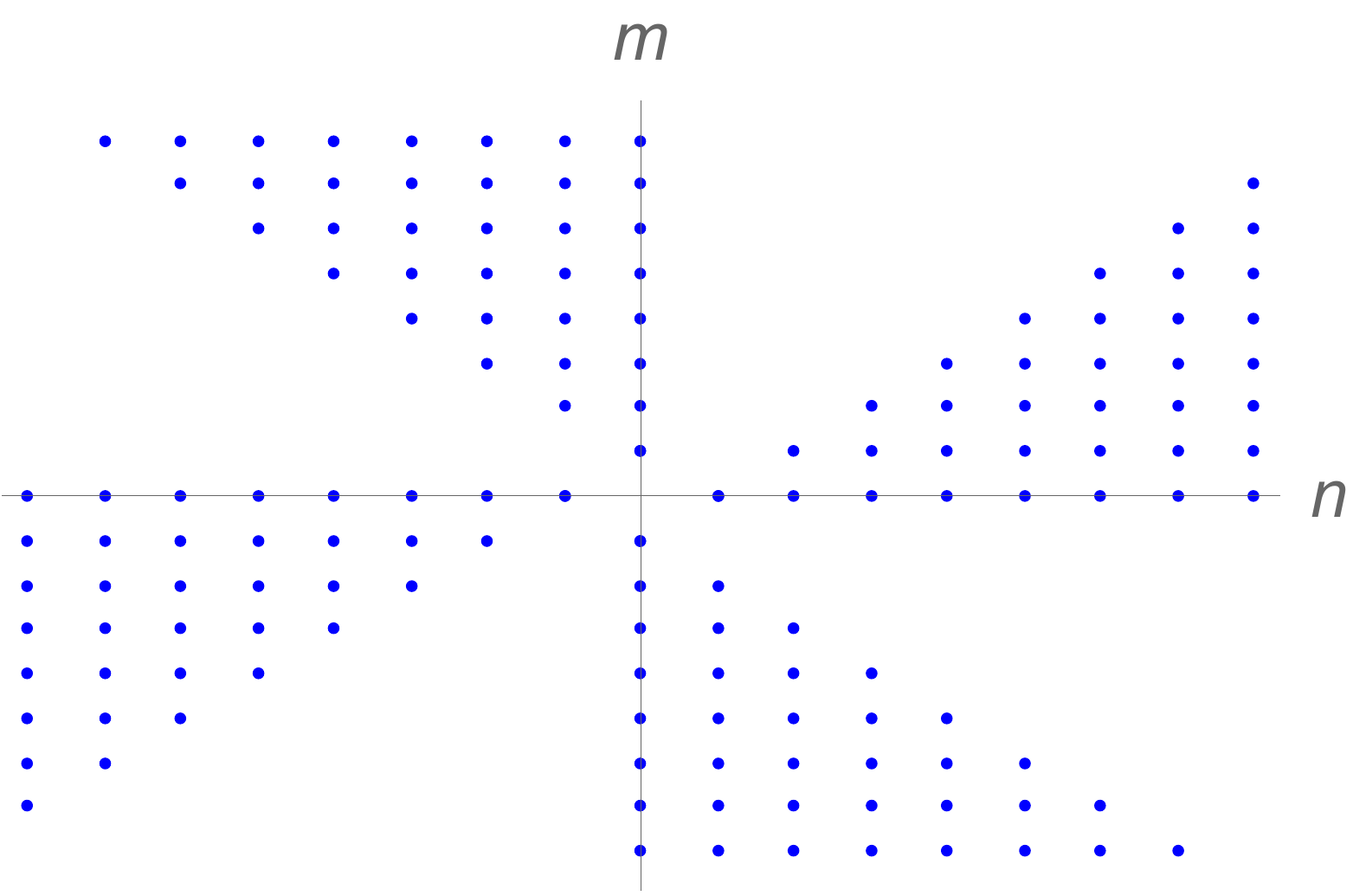}
\caption{ The included quadrants are: region i) $n\geq 1,  0\leq m \leq n{-}1$, region ii) $m\geq 1,   -m{+}1 \leq n \leq 0$,  region iii) $n\leq{ -}1,  1{-}|n|\leq m \leq 0$, and region iv) $m\leq -1,   0\leq n \leq |m|{-}1$~.} \label{figQ}
\end{figure}
We have used the Fourier basis $e^{i (m x_0 + n p_0)}$, with $m,n$ running over the integers,  to represent some function $f(x_0,p_0)$, see Eq.~\ref{fourier}. Crawford and Cary \cite{CrawfordCary} (see also \cite{CaryMeiss, MEISS1983375}) also use the Fourier basis, but they label the points differently: $(m, n)$ are restricted to quadrants as shown in Fig.~\ref{figQ}, and there is an additional integer index $\al$: instead of $e^{2\pi i (m x_0 + n p_0)}$ one has,
\be\label{psial}
\psi_{\al, (m, n)}(x_0, p_0) =e^{ 2\pi i (m x_{\al} + n p_{\al})} = \exp\[ 2\pi i \Big( (m( 2 d_{\al} {-} d_{\al-1}) + n d_{\al}) x_0 + ( n(2d_{\al-1} {-} d_{\al-2}) + m d_{\al})p_0\Big)\]
\ee
where $x_{\al}$ and $p_{\al}$ are the time evolution of $x_0$, and $p_0$, as in (\ref{xpt}). The expansion of $g(x_0, p_0)$ is thus, 
\be \label{418}
g(x_0, p_0) = \!\!\sum_{\al=-\infty}^{\infty} \sum_{m,n=-\infty}^{\infty} g_{\al, (m, n)} \ \psi_{\al, (m,n)}(x_0,p_0)~,
\ee
where $g_{\al, (m, n)}$ is only nonzero if $(m,n)$ lie in one of the four quadrants shown in Fig.~\ref{figQ}. The advantage of this labeling of the points is that under time evolution, $\psi_{\al, (m,n)}(x_0, p_0) \rightarrow \psi_{\al, (m,n)}(x_t, p_t) =\psi_{\al+t, (m,n)}(x_0, p_0) $.
 Expanding $f(x_t, p_t)$ in this basis, in the same way as we expanded $g(x_0, p_0)$ but with $x_t$ and $p_t$ appearing instead, 
 \be
 f(x_t, p_t) = \!\!\sum_{\al=-\infty}^{\infty}\sum_{m,n} f_{\al, (m,n)} \ \psi_{\al, (m,n)}(x_t,p_t)~.
\ee
Shifting $\al \rightarrow \al-t$  we get
 \be \label{420}
 f(x_t, p_t) = \!\!\sum_{\al=-\infty}^{\infty}\sum_{m,n} f_{\al-t,(m,n)} \ \psi_{\al,(m,n)}(x_0,p_0)~.
\ee
The two-point function is thus, 
\be \label{2CC}
\la f(x_t, p_t) g^*(x_0, p_0)\ra  =\sum_{\al=-\infty}^{\infty}  \sum_{m,n}\, f_{\al-t, (m,n)}\ g^*_{\al,(m,n)}~,
\ee
where we used that $\la \psi_{\al, (m,n)} \psi^*_{\beta, (a, b)} \ra = \delta_{\al, \beta} \delta_{m, a} \delta_{n, b}$.

This basis is used in  \cite{CrawfordCary} to prove  that if the function $f(x,p)$ is sufficiently smooth (so that all partial derivatives up to  the $r$'th order derivative  are square integrable), then the two-point function of $f$ decays at least as fast as the $r$'th Ruelle resonance.
To see this, from (\ref{420}) one notices that each derivative of $f(x_t,p_t)$ brings down a factor that scales as $\lam^{\al}$ inside the sum. Thus, if one takes $r$ derivatives, in order for the sum to converge the coefficients $f_{\al, (m,n)}$ must decay at least as fast as $\lam^{- |\al| r}$ and correspondingly the two-point function (\ref{2CC}) decays at least as fast as $\lam^{- |\al| r t}$ . 

There is an intuitively satisfactory explanation for the decay of the connected correlation function: under time evolution, small wavenumber modes evolve into large wavenumber modes. Therefore, if the function $f$ is comprised of only small wavenumber modes, its connected two-point function will rapidly give zero. Heuristically, the more smooth the function $f$ is, the fewer large wavenumber modes comprise it, consistent with the result that if the $r$'th derivative is smooth, then the two-point function decays with at least the $r$'th Ruelle resonance.

The decay of the two-point function we found in (\ref{415})  is consistent with the continuity result of \cite{CrawfordCary}. For our $f(x,p) = B_k(x) B _q(p)$, one can take $k+q$ derivatives of it before getting zero on the next additional derivative. The expression for the two-point function (\ref{2ptBB}) can of course be found in this basis, by using (\ref{418}) in order to express $e^{2\pi i (m x_0 + n p_0)}$ in terms of $\psi_{\al, (m,n)}(x_0, p_0)$, but this is not useful. \\

\noindent \textit{Functions with a simple two-point function}

The utility of this alternate basis is to answer a different question: which functions have a two-point function that has simple time behavior? In other words, which functions have a two-point function that decays precisely as a single Ruelle resonance, rather than just decaying asymptotically at late times as a Ruelle resonance.

To get simple behavior in time for the two-point function $\la g(x_t, p_t) g(x_0, p_0)\ra$, we take the coefficients $g_{\al, (m,n)}$ to be seperable in their $\al$ and $(m,n)$ dependence and to take the form, 
\be \label{ffac}
g_{\al, (m,n) } =e^{- |\al| r \log \lam} g_{(m,n)}
\ee
where $r$ is some positive integer and $g_{(m,n)}$ is arbitrary. The two-point function (\ref{2CC}) becomes, 
\bea
\la g(x_t, p_t) g(x_0, p_0)\ra\!\! \!\! &=&\sum_{m,n} \,|g_{(m,n)}|^2 \!\sum_{\al=-\infty}^{\infty} \!\!e^{- |\al -t | r \log \lam} e^{- |\al | r \log \lam} \\ \nn
 &=&\(\frac{2}{1- \lam^{-2r}} + (t{-}1)\)e^{-t\, r \log \lam}\sum_{m,n} \,|g_{(m,n)}|^2 ~,
\eea
where to get the second line we evaluated the sum over $\al$ appearing in the first line. As promised, the time dependence is simple.

\sss*{Limitations of shadowing}
The equation of motion  (\ref{24}) for the cat map may be written as, 
\be
\delta_t^2 x_t = (s-2) x_t - m_{t+1}~,
\ee
where $m_{t+1}$ is the integer necessary to ensure that $x_{t+1}$ is in the range $0$ to $1$,  taking on the values  $m_{t+1} \in \{-1, 0,1, \ldots s{-}1 \}$. As we did for the Bernoulli map (see Eq.~\ref{xt}), we  regard $m_{t+1}$ as a source for a linear map. The solution of the equations of motion is thus \cite{GC}, 
\be\label{xtc}
x_t = \sum_{t' = - \infty}^{\infty} g_{t t'} m_{t'}~, \ \ \ \ \ \ g_{t t'} =\frac{\lam^{-|t-t'|}}{\lam- \lam^{-1}}~.
\ee
As in the Bernoulli map, correlation functions of $x_t$ can be expressed in terms of correlation functions of $m_t$, see Eq.~\ref{321}. Let us start by checking the one-point function, $\la x_t\ra$. We first compute $\la m_t\ra$,
\be
\la m\ra =\sum_m P(m) m = \frac{1}{s}\sum_{m=0}^{s-2} m - \frac{1}{2s} + \frac{s-1}{2s}= \frac{s-2}{2}~,
\ee
where we used the measure $P(m) = \frac{1}{2s}$ for $m=-1$ and $m=s-1$, and  $P(m) = \frac{1}{s}$ for the other $m$ \cite{GC}.
As a result, using (\ref{xtc}) we see that, 
\be
\la x_t\ra = \sum_{t'=-\infty}^{\infty} g_{t t'} \la m_{t'}\ra = \frac{1}{\lam-\lam^{-1}} \sum_{t'=-\infty}^{\infty} \lam^{-|t-t'|} \frac{s-2}{2} = \frac{1}{2}~,
\ee
which is the correct answer. 
Because $g_{t t'}$ decays exponentially with $|t-t'|$, one gets approximately the right answer by restricting $t'$ to be close to $t$ (within an order-one integer $l$), 
\be
\la x_t\ra \approx \sum_{t'= t-l}^{t' = t+l} g_{t t'} \la m_{t'}\ra~,
\ee
up to corrections of order $\lam^{- l-1 }$. 
This is referred to as shadowing in \cite{GC}.

 Consider now a two-point function, such as e.g. $\la x_{t} p_t x_0 p_0\ra$. Expressing $p_t$ as $p_t = x_{t} - x_{t-1}$, this will involve correlation functions of the type, e.g. $\la x_0 x_1 x_{t-1} x_t\ra$. Via (\ref{321}), to compute this we need to know  $\la m_{t_1'} \cdots m_{t_4'}\ra$. As a result of the exponential decay of the Green's function, we only need $\la m_{t_1'} \cdots m_{t_4'}\ra$ for $t_i'$ that are approximately in the range between $-l$ and $t+l$, where $l$ is an order-one number. This will be enough to get the approximate correlation function of the $x$'s, up to a correction of order $\lam^{- l-1 }$. Computing the correlation function of the $m$'s requires knowing the measure $P(m_{-l} m_{-l+1} \cdots m_{t+l})$ for blocks of length $t + 2l$. In \cite{GC}, an algorithm was given to find this measure. The complexity increases with the size of the block. So this is not an efficient way of computing correlation functions with large time separation $t$.

\section{Spatiotemporal cat map} \label{sec:4}
\begin{figure}
\centering
\includegraphics[width=6in]{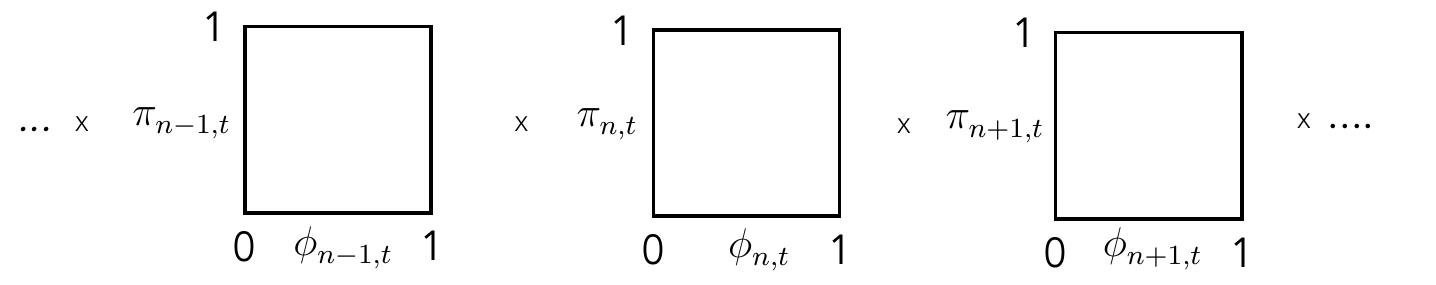}
\caption{The spatiotemporal cat map acts on a product of tori (over site $n$) with phase space $(\phi_n, \pi_n)$ at site $n$. } \label{phifig}
\end{figure}
The spatiotemporal cat map \cite{Gutkin16} acts on a product of tori, $T^2 \times T^2 \times \cdots \times T^2$. The phase space is a product over $n$ of $(\phi_n, \pi_n)$ and there is locality in $n$, see Fig~\ref{phifig}. The equations of motion are \cite{GC}, 
 \be \label{sceom}
 \( \delta_t^2 + \delta_n^2\) \phi_{n,t} = (s-4) \phi_{n,t}~\text{ mod } 1~,
 \ee
 where $s$ is an integer and the definition of $\delta^2$ is the same as it was for the cat map (\ref{24}): $\delta_t^2 \phi_{n,t} = \phi_{n,t+1} - 2 \phi_{n,t} + \phi_{n, t-1}$ and analogously, $\delta_n^2 \phi_{n,t} = \phi_{n+1, t} - 2\phi_{n,t} + \phi_{n-1, t}$. The spatiotemporal cat map is the natural spacetime generalization of the cat map, with symmetry between the spatial ($n$) direction and the time ($t$) direction.  It can be viewed as a discretization of a free scalar field  (having equations of motion $(-\d_t^2 + \d_x^2) \phi =m^2 \phi$) with a negative mass squared, $m^2 = -(s{-}4)$,  and a spatial gradient term of the wrong sign. We focus on $s>4$, for which the map is chaotic. 
 The equations of motion can just as well be written as an action on phase space, using $\pi_{n, t} = \phi_{n, t} - \phi_{n, t-1}$, this gives \cite{Gutkin16}
\bea
\phi_{n, t+1} &=& - \phi_{n-1, t} - \phi_{n+1, t} + \pi_{n, t} + (s-1) \phi_{n, t} \ \text{  mod } 1 \\
\pi_{n, t+1} &=&  - \phi_{n-1, t} - \phi_{n+1, t} + \pi_{n, t} + (s-2) \phi_{n, t}\  \text{  mod } 1 ~.
\eea
\sss*{Solving the equations of motion}
Let us now solve the equations of motion (\ref{sceom}). We will find that in Fourier space for the spatial ($n$) direction, the spatiotemporal cat map is a collection of cat maps. 
By inspection the solution of (\ref{sceom}) is, 
 \be
 \phi_{n, t} = \lam^t \rho^n~, \ \ \ \ \ \lam + \frac{1}{\lam} + \rho + \frac{1}{\rho} = s~,
 \ee
 where $\rho$ is still to be determined. 
 Assuming there are $N$ sites, and placing periodic boundary conditions, requires $\rho^N = 1$. This gives $\rho = \rho_k$ and $\lam = \lam_k$ where
 \be \label{1916}
 \rho_k = e^{2 \pi  i k/N}~, \  \ \ \ \ \ \lam_k + \frac{1}{\lam_k} + 2\cos\(\frac{2\pi k}{N}\) = s~.
 \ee
We can therefore expand the field in terms of these solutions, 
\be
\phi_{n, t} = \sum_{k=0}^{N-1}  \rho_k^n \( c_k \lam_k^t + d_k \lam_k^{-t} \)~,
\ee 
where $\lam_k$ is given by (\ref{1916}) (either one of the two solutions) and $c_k$ and $d_k$ are arbitrary and set by initial conditions, 
\be
c_k + d_k = \frac{1}{N} \sum_{n=0}^{N-1} \phi_{n,0} e^{- 2\pi i n k/N} \equiv \t \phi_{k,0}~,
\ee
where $\t \phi_{k,0}$ indicates the Fourier transform. Defining $\t \pi_{k,0}$ analogously, $ \t \pi_{k,0}\equiv \frac{1}{N} \sum_{n=0}^{N-1} \pi_{n,0} e^{- 2\pi i n k/N}$,
and solving for $c_k$ and $d_k$ in terms of initial conditions, we have that the time evolution of the Fourier mode, $
\t \phi_{k,t} = c_k \lam_k^t  + d_k \lam_k^{-t}
$, is 
\bea \label{2023}
\t \phi_{k,t} &=& \frac{\t \phi_{k,0}}{1 +\lam_k} \( \lam_k^{t+1} +\lam_k^{-t}\) + \frac{\t \pi_{k,0}}{1- \lam_k^2} \lam_k \(- \lam_k^t + \lam_k^{-t}\)\\
\t \pi_{k,t}&=& \t \phi_{k,0} \frac{1- \lam_k}{1+ \lam_k} \( -\lam_k^t + \lam_k^{-t}\) + \frac{\t \pi_{k,0}}{1+\lam_k} \(\lam_k^t + \lam_k^{1-t}\)~.
\eea
Comparing with the cat map, we see that each $k$ mode is like a cat map, with an effective $s$ that is $s - 2\cos\(\frac{2\pi k}{N}\)$. \\

\begin{figure}
\centering
\includegraphics[width=2.2in]{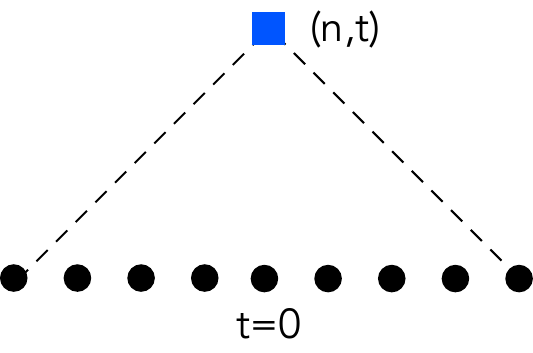}
\caption{The field at site $n$ and time $t$, $\phi_{n,t}$, can be expressed in terms of the fields $\phi_{n',0}$ and $\pi_{n',0}$ at time $t=0$ which are in the past lightcone, see (\ref{610}).} \label{Green}
\end{figure}

There is an alternative way of writing $\phi_{n,t}$, in terms of a retarded Green's function, that will be more useful to us. By causality, $\phi_{n,t}$ is determined by the $\phi_{n', 0}$ and $\pi_{n',0}$ that lie within the past lightcone of $(n,t)$: $n'$ in the range $n-t\leq n'\leq n+t$, see Fig.~\ref{Green}, 
\be \label{610}
\phi_{n, t} = \sum_{n'=n-t}^{n+t} \(g_{|n-n'|,t} \, \phi_{n',0} + h_{|n-n'|, t}\, \pi_{n',0}\)~.
\ee
For instance, for $\Delta t=2$, through explicit use of the equations of motion, 
\be
\phi_{n, t+2} = \(\phi_{n-2, t}{+}\phi_{n+2,t}\) +(1-2s) \( \phi_{n-1,t} {+} \phi_{n+1,t}\)  + (s^2-s+1) \phi_{n, t}
- \( \pi_{n-1,t} {+} \pi_{n+1,t}\) + s\, \pi_{n, t}~.
\ee
We may compute  the retarded Green's function in order to explicitly get $g_{|n-n'|,t}$ and $h_{|n-n'|, t}$ for general $t$, but this won't be necessary.

\sss*{Vanishing local connected correlation functions}
There is a broad range of functions of $\phi_n, \pi_n$ (``operators'') for which we can compute correlation functions. We will focus on local operators, localized at site $n$ at time $t$, 
\be \label{mF}
\mF_{n,t} = \sum_{r, s=-\infty}^{\infty} f_{r, s}\, e^{2\pi i (r \phi_{n,t} + s \pi_{n,t})}~,
\ee
where $f_{r,s}$ are arbitrary coefficients. This is the most general local operator, satisfying periodicity $\phi\sim \phi +1$ and $\pi\sim \pi+1$. Let us consider the two-point function between $\mF_{n,t}$ and another local operator, $\mathcal{G}_{m,0}$, localized at site $m$ at time $0$, $\la \mF_{n,t}\, \mathcal{G}_{m,0}\ra$ where, 
\be \label{mG}
\mathcal{G}_{m,0} = \sum_{a, b=-\infty}^{\infty} g_{a, b}\, e^{2\pi i (a \phi_{m,0} + b \pi_{m,0})}~.
\ee
The correlation function is defined in the same way as it was for the cat map, as an integral over all initial conditions $\phi_{n,0}, \pi_{n,0}$, 
\be
\la \mF_{n,t}\, \mathcal{G}_{m,0}\ra = \int \prod_n d\phi_{n,0} d\pi_{n,0}\ \mF_{n,t}\, \mathcal{G}_{m,0}
\ee
Inserting $\mF_{n,t}$ in (\ref{mF}) and $\mathcal{G}_{m,0}$ in (\ref{mG}), 
\be
\la \mF_{n,t}\, \mathcal{G}_{m,0}\ra  =  \sum_{r, s, a,b=-\infty}^{\infty} f_{r, s} g_{a, b}\, \la  e^{2\pi i (r \phi_{n,t} + s \pi_{n,t})}e^{2\pi i (a \phi_{m,0} + b \pi_{m,0})}\ra~.
\ee
To compute this correlation function, we need to express $\phi_{n,t}$ and $\pi_{n,t}$ in terms of $\phi_{n',0}$ and $\pi_{n',0}$. To do this, we make use of (\ref{610}). We also write $\pi_{n,t} = \phi_{n,t} - \phi_{n,t-1}$. We get, 
\be \label{616}
r \phi_{n,t} + s \pi_{n,t} = \sum_{n'=n-t}^{n+t} \(\big( (r{+}s) g_{|n{-}n'|, t} -s g_{|n-n'|,t{-}1}\big) \phi_{n',0} +\big((r{+}s) h_{|n{-}n'|, t} -s h_{|n-n'|,t{-}1}\big) \pi_{n',0} \)~.
\ee
Let assume the time $t\geq 3$. Consider two different $n'$ in the sum  (\ref{616}), $n' = n_1$ and $n'=n_2$, such that $n_1, n_2 \neq m$. Each of these terms must vanish in order to give a nontrivial contribution after integration. This in particular implies that $
(r{+}s) g_{|n{-}n_i|, t}  = s g_{|n-n_i|,t{-}1}$  for $i=1,2$ and, thus, 
\be
r = s \frac{ g_{|n{-}n_1|, t} - g_{|n-n_1|,t{-}1}}{g_{|n{-}n_1|, t}}=s \frac{ g_{|n{-}n_2|, t} - g_{|n-n_2|,t{-}1}}{g_{|n{-}n_2|, t}}~.
\ee
For this to be satisfied we  must have $s=r=0$. Therefore, only $s=r=0$ contributes and the correlation function factorizes, 
\be
\la \mF_{n,t}\, \mathcal{G}_{m,0}\ra  = \la \mF_{n,t}\ra\la \mathcal{G}_{m,0}\ra~,
\ee
and the connected two-point function, $\la \mF_{n,t}\, \mathcal{G}_{m,0}\ra  -\la \mF_{n,t}\ra\la \mathcal{G}_{m,0}\ra$, vanishes. 

It is clear that adding more local operators at $t=0$ would not alter this conclusion: the correlation function 
\be
 \la  e^{2\pi i (r \phi_{n,t} + s \pi_{n,t})}\prod_{i=1}^k e^{2\pi i (a_i \phi_{m_i,0} + b_i \pi_{m_i,0})}\ra
 \ee
  again factorizes into one point function, unless we take $k$ to be $2t-1$, and have all the $m_i$ distinct and within the past lightcone of $(n,t)$: $n-t\leq m_i\leq n+t$. 
  
  Likewise, we believe that correlation functions of multiple local operators at different times, such as e.g. a three-point function, $\la \mF_{n_1,t_1}  \mathcal{H}_{n_2,t_2} \mathcal{G}_{m,0}\ra$ will also factorize into one-point functions. \\
  
 \noindent \textit{Geometric interpretation}
 
  \vspace{.1cm}
For ergodic systems we expect connected correlation functions to decay exponentially rapidly with time. Their vanishing - for local operators - for the spatiotemporal cat map would appear to be indicative of the high degree of ergodicity of the map. Let us give a geometric interpretation of this effect. 
  
 Let us actually first return  to the cat map. From (\ref{twopC}) we see that the connected two-point function of a function that depends only on $x$ (or only on $p$) vanishes, 
 \be \label{2ptz}
 \la f(x_t) f(x_0)\ra - \la f(x_t) \ra \la f(x_0)\ra = 0~,
 \ee
 which follows from (\ref{twopC}) because if $f(x, p)$ has no $p$ dependence, $n=0$ and $b=0$, and so $\la f(x_t) f(x_0)\ra = f_{0,0} f_{0,0}$.  Geometrically, the two-point function $\la f(x_t,p_t) f(x_0,p_0)\ra$ takes some distribution $f(x,p)$ in $(x,p)$ phase space, evolves it forward in time to time $t$, and computes the overlap with the initial distribution. Consider an $f(x,p)$ that depends only on $x$ -- for instance, an $f$ that is localized to a strip in the $x$ direction while constant in the $p$ direction. To have $\la f\ra =0$ we can, for instance, take $f(x)$ to be $1$ in the upper half of the strip and $-1$ in the lower half. The result (\ref{2ptz}) means that if we evolve this distribution forward in time, then it has no overlap with the initial distribution. 
 
 Turning now to the spatiotemporal cat map, the phase space  $\prod_n (\phi_n, \pi_n)$ is $2N$ dimensional (if there are $N$ sites). Our result is that for the connected two-point function of a function to be nonzero at time $t$, the function must depend on at least $t$ directions in phase space (in particular, the $t$ directions have to come from $t$ different sites).  Therefore, to obtain a nonzero connected two-point function at late times, one needs to be computing the two-point function of a function that is highly nonlocal. 
 In a physical measurement, one expects to only measure a few variables while averaging over the others -- in other words, one would consider correlation functions of a function that is uniform in most directions in phase space. This result indicates that such a connected correlation function immediately vanishes: the spatiotemporal cat map is highly ergodic. 
 
 This effect is due entirely to the linearity of the map. If one were to add a nonlinear term to the spatiotemporal cat map, our equation (\ref{610}) wouldn't apply: the field at time $t$ would not simply be the field at time zero (it would also involve higher powers of the field). Correspondingly, it would no longer be the case that connected correlation functions of local operators vanish.

\section{Discussion}\label{sec:5}
While the initial motivation for chaotic dynamics arose from many-body systems, much of the early discussion focused on few-body systems. Textbook treatments of chaos continue to focus on few-body systems. Within an eye towards many-body chaos, we have looked at correlation functions in some of the simplest chaotic maps. The linearity of these maps provides analytic tractability. 

A natural next step is to develop techniques for computing correlation functions for maps with nonlinearity, such as those in \cite{LiangCvit}. It would also be good to connect the classical spatiotemporal cat map with recent studies of models of quantum many-body chaos. It has been argued \cite{Gutkin:2020hro,Gutkin:2020_2} that the spatiotemporal cat map is related to the dual unitary model \cite{BertiniKosProsen2019,BertiniKosProsen2019_2,CiracProsen:2019umh,ClaeysLamacraft2020,GopalakrishnanLamacraft2019,ReidBertini,ZhouNahum2020}, closely related to random unitary circuits, \cite{NahumVijayHaah,PSQ:2020,BertiniPiroli2020,ZhouNahum2019,BaoChoiAltman,FanVijayVishwanathYou}. A prominent recent model of quantum many-body chaos is the SYK model \cite{Kitaev,SY, PR, Maldacena:2016hyu,Rosenhaus:2018dtp}, a model of a large number $N$ of fermions with random all-to-all interactions, with  correlation functions which are analytically computable in a $1/N$ expansion \cite{Gross:2017aos}. In this paper  we discussed linear, area-persevering, maps on the torus and products of tori. The same Fourier space techniques used to compute the correlation functions can be applied to generalizations of the cat map acting on other compact symmetric spaces, such as \cite{Nance2011,ChenMaoChui2004}. 

An ambitious long-term goal is developing techniques to compute correlation functions in chaotic classical and quantum many-body systems. The expected rapid decay in space and time of correlation functions differs strongly from the oscillatory behavior familiar in free theories. In the context of the cat maps studied here, the decay of a two-point function has a clear intuitive explanation: the low wavenumber Fourier modes comprising the function rapidly evolve into high wavenumber  modes, which have little support in the function. One might hope this insight will find use in understanding the decay of correlation functions in more realistic systems.

\sss*{Acknowledgments} 
We thank Predrag Cvitanovi\'c, Boris Gutkin, and Enrique Pujals for helpful discussions. X.-Y. H. is supported by the James Arthur Graduate Award.

\bibliographystyle{utphys}

\end{document}